\documentclass[a4paper, conference, 10pt]{IEEEtran}
\IEEEoverridecommandlockouts
\usepackage{acronym}
\usepackage[left=1.57cm,right=1.57cm,top=0.95cm,bottom=2.54cm]{geometry}

\usepackage{cite}
\usepackage{amsmath,amssymb,amsfonts}
\usepackage{graphicx}
\usepackage{textcomp}
\usepackage{xcolor}
\def\BibTeX{{\rm B\kern-.05em{\sc i\kern-.025em b}\kern-.08em
    T\kern-.1667em\lower.7ex\hbox{E}\kern-.125emX}}

\usepackage{url}
\usepackage{bbm}
\usepackage[font=small,skip=0pt]{caption}

\usepackage[utf8]{inputenc}
\usepackage{algorithm}
\usepackage[noend]{algpseudocode}
\algnewcommand{\Initialize}[1]{%
  \State \textbf{Initialize:}
  \Statex \hspace*{\algorithmicindent}\parbox[t]{.8\linewidth}{\raggedright #1}
}

\DeclareMathOperator*\argmax{arg \, max}		
\DeclareMathOperator*\argmin{arg \, min}		
\DeclareMathOperator*\maximize{maximize}

\newcommand{\field}[1]{\mathbb{#1}}

\newcommand{\set}[1]{\mathcal{#1}}

\newcommand{\R}{{\field{R}}}
\newcommand{\N}{{\field{N}}}
\newcommand{\Expt}{{\field{E}}}

\newcommand{\ve}[1]{\boldsymbol{\mathbf{#1}}}

\newcommand{\vs}{\ve{s}}
\newcommand{\va}{\ve{a}}

\newcommand{\vx}{\ve{x}}

\newcommand{\Ns}{\set{N}}
\newcommand{\Ss}{\set{S}}
\newcommand{\Ks}{\set{K}}
\newcommand{\Ds}{\set{D}}
\newcommand{\As}{\set{A}}
\newcommand{\Bs}{\set{B}}
\newcommand{\Es}{\set{E}}

\newcommand{\Os}{\set{O}}
\newcommand{\Ts}{\set{T}}

\newcommand{\operator}[1]{\mathrm{#1}}

\newcommand{\HF}{\operator{(HF)}}
\newcommand{\PP}{\operator{(HOPP)}}
\newcommand{\Pp}{\operator{(PP)}}
\newcommand{\LHO}{\operator{(HOL)}}
\newcommand{\EHO}{\operator{(HOE)}}
\newcommand{\WHO}{\operator{(HOW)}}
\newcommand{\ATT}{\operator{(HOA)}}
\newcommand{\SHO}{\operator{(HOS)}}
\newcommand{\TS}{\operator{(T)}}
\newcommand{\LS}{\operator{(D)}}

\newcommand{\off}{\operator{(off)}}
\newcommand{\on}{\operator{(on)}}
\newcommand{\Act}{\operator{(A)}}
\newcommand{\Eg}{\operator{(E)}}

\begin{document}

\title{Knowledge Transfer in Deep Reinforcement Learning for Slice-Aware Mobility Robustness Optimization}

\author{
\IEEEauthorblockN{Qi Liao\IEEEauthorrefmark{1}, Tianlun Hu\IEEEauthorrefmark{1}\IEEEauthorrefmark{2}, Dan Wellington\IEEEauthorrefmark{3}}
\IEEEauthorblockA{ 
\IEEEauthorrefmark{1}Nokia Bell Labs, Stuttgart, Germany\\
\IEEEauthorrefmark{2}Department of Informatics, Technical University of Munich, Germany\\
\IEEEauthorrefmark{3}Nokia Software, Bellevue, United States\\
Email: \url{qi.liao@nokia-bell-labs.com}, {\{tianlun.hu, dan.wellington\}@nokia.com}}
}

\maketitle

\begin{abstract}
The legacy \ac{MRO} in self-organizing networks aims at improving handover performance by optimizing cell-specific handover parameters. However, such solutions cannot satisfy the needs of next-generation network with network slicing, because it only guarantees the received signal strength but not the per-slice service quality. To provide the truly seamless mobility service, we propose a deep reinforcement learning-based \ac{SAMRO} approach, which improves handover performance with per-slice service assurance by optimizing slice-specific handover parameters. Moreover, to allow safe and sample efficient online training, we develop a two-step transfer learning scheme: 1) regularized offline reinforcement learning,  and 2) effective online fine-tuning with mixed experience replay. System-level simulations show that compared against the legacy \ac{MRO} algorithms, \ac{SAMRO} significantly improves slice-aware service continuation while optimizing the handover performance. 
\end{abstract}

%
\section{Introduction}
The concept of \ac{SON} was introduced nearly a decade ago when various use cases of \ac{SON} were standardized by 3GPP \cite{TS32500}. However, the introduction of network slicing into 5G and beyond raises new challenges to \ac{SON}, while the slice-aware \ac{SON} has rarely been studied. 
Slice-aware \ac{SON} requires solutions that satisfy the {\it slice-specific} service requirements, while the legacy \ac{SON} solutions optimize {\it cell-specific} or {\it cell-pair-specific} (dedicated with respect to a neighboring cell) parameters by observing cell-specific \acp{KPI}.

In this paper, we study a particular use case of slice-aware \ac{SON}, namely, the slice-aware mobility robustness optimization (SAMRO). The legacy \ac{MRO} has the objective of minimizing the \ac{HO}-related \acp{RLF} and the number of unnecessary handovers. The decision variables are a set of cell-specific and cell-pair-specific \ac{HO} parameters. The complexity and dynamics of user mobility result in a \ac{HO} performance metric-related objective function that is in general non-convex, non-linear, and dependent on a high-dimensional \ac{HO} parameter space. Thus, practical solutions include heuristics \cite{nguyen2017mobility}, reinforcement learning \cite{mwanje2014distributed}, and Bayesian methods \cite{liao2013statistical}. However, these solutions, aiming to optimize the signal quality-based \ac{HO} performance metrics, cannot guarantee the per-slice service continuation in terms of throughput and latency. A recent work \cite{9405369} proposed to improve the quality of experience while optimizing the \ac{HO} performance. However, with the cell-specific parameters, the \ac{HO} optimization cannot be customized for every slice. 

We propose a \ac{DRL}-based solution that provides a more seamless mobility service through the introduction of slice-specific \ac{HO} parameters and performance metrics. Although the newly introduced parameters increase the dimensions of the state and action spaces, by utilizing model-free actor-critic architectures such as \ac{TD3} \cite{fujimoto2018addressing}, we can work with large continuous state and action spaces with significantly reduced model complexity, compared to the value-based algorithms such as \ac{DQN} \cite{mnih2015human}. 
 
Even with the help of the actor-critic architecture, we still face two challenges when developing a practical solution:  
\begin{itemize}
\item {\it Large discrete action space}: In practical systems, \ac{HO} parameters are selected from finite discrete sets, but directly working with huge discrete action space leads to intractable complexity. How to project the continuous agent actions to the discrete operating actions?  
\item {\it Effective training with knowledge transfer}: \ac{DRL} approaches usually require substantial exploration, while in practice online training with unconstrained exploration is expensive. On the other hand, a large amount of biased data -- usually collected in the safe operating space -- is available.  
How to train more effectively by transferring knowledge from the biased offline data?
\end{itemize}
%
To overcome these challenges, we propose a two-step transfer learning-assisted \ac{SAMRO} scheme, consisting of regularized offline \ac{DRL} model training and online fine-tuning. The main contribution is threefold:  
\begin{itemize} 
\item {\it Mapping between continuous and discrete actions}: Motivated by \cite{dulac2015deep}, we use approximate nearest-neighbor methods to find the mapping between the continuous agent actions and the discrete operating actions. 
\item {\it Regularized offline \ac{DRL} training}: We develop a regularized offline \ac{DRL} algorithm to exploit the offline data, while dealing with the bias by regularizing the objective with the density estimates of state-action pairs. 
\item {\it Sample-efficient online fine-tuning}: We propose mixed experience replay during the online fine-tuning. The newly collected samples are mixed with a selected subset of offline samples for the batch training. 
\end{itemize}
%
The rest of the paper is organized as follows. In Section \ref{sec:Model} we define the system model and the slice-specific \ac{HO} parameters and metrics. The \ac{SAMRO} problem is discussed in Section \ref{sec:Problem}. Section \ref{sec:Solution} introduces the proposed transfer learning and actor-critic-based solution. The numerical results are shown in Section \ref{sec:Numerical}. Section \ref{sec:Conclusion} summarizes the conclusions.


\section{System Model}\label{sec:Model}
We consider a set of cells in the neighboring area, denoted by $\Ns$. Each cell can support numerous instances of different slices. 
Let the set of slices be denoted by $\Ss$. A service of slice $s\in\Ss$ has a defined throughput requirement $\eta^{\ast}_s$ and latency requirement $d^{\ast}_s$.
Assume that there are a set of $B$ directional cell boundaries, denoted by $\Bs:=\left\{(n_i,m_i): n_i, m_i\in\Ns; i = 1, \ldots, B\right\}$. In practice, if the boundary $(n, m)\in\Bs$, it means that cell $m$ is within the neighboring cell list of cell $n$. Note that the boundary is directional, i.e., $(n, m) \neq (m, n)$ for $m\neq n$. Assume that if $(n, m)\in\Bs$, then $(m, n)\in\Bs$. Therefore, there are $B/2$ combinations (regardless of the order) of the neighboring cell-pairs. We denote the set of cell-pairs as $\Bs'$ with $|\Bs'|=B/2$. 

\subsection{\ac{HO} Process and Slice-Specific Parameters}\label{subsec:HO_Params}
%
In the legacy \ac{MRO}, for any user served by cell $n\in\Ns$, a simplified \ac{HO} criterion\footnote{The complete version can be found in \cite[Section 5.5.4.4]{TS38331}, but many offsets are not needed for intra-frequency handover.} for a handover from cell $n$ to cell $m$ when $(n,m)\in\Bs$ is that
$P_m(t) > P_n(t) + O_{n, m} $ holds for a time duration of $T_{n, m}$, where $P_m$ and $P_n$ (in dBm) indicates the received signal strength from cell $m$ and $n$ respectively, and $O_{n, m}$ (in dB) and $T_{n, m}$ (in ms) are the \ac{HOM} and \ac{TTT} from cell $n$ to cell $m$. 

The \ac{HOM} and \ac{TTT} defined above are on the cell-pair basis. We propose to define them also on the slice basis, such that the \ac{HO} decision also considers whether the neighboring cell can provide the required service performance. The modified \ac{HO} criterion for a service in slice $s$ from cell $n$ to cell $m$ at time $t'$ (in ms) is given by
\begin{equation}
P_m(t)  > P_n(t) + O_{n,m,s},  \ \forall t \in [t'-T_{n,m,s}+1, t'].  
\label{eqn:Slice_HO}
\end{equation} 
Note that without loss of generality, \eqref{eqn:Slice_HO} can be generalized to mobility-specific or even user-specific \ac{HO} criteria. For example, we can group the users based on both mobility and slice classes, and define the user group-specific \ac{HO} parameters. Since mobility-specific handover has been well studied \cite{tesema2016evaluation}, in this paper we focus on the slice-specific aspect to ensure the service continuation.

\subsection{Slice-Specific \ac{HO} Performance Metrics}\label{subsec:HOMetrics}
The conventional \ac{HO} optimization aims to reduce the following four \ac{HO} events:
\begin{itemize}
\item {\it \ac{HOL}}: When a user is leaving the coverage area of its serving cell $n$ towards the target cell $m$, but the handover is triggered too late which causes \ac{RLF} before completing a handover.   
\item {\it \ac{HOE}}: This occurs when the \ac{HO} decision is made too early. The neighboring cell cannot provide a sustainable signal quality to the user and a \ac{RLF} happens right after the handover.
\item {\it \ac{HOW}}: This happens when the user is handed over to a wrong cell. A \ac{RLF} is detected shortly after a successful handover to the target cell and then the user is connected to another neighboring cell. 
\item {\it \ac{HOPP}}: When a user is handed over from cell $n$ to cell $m$, but after a short time period the handover from cell $m$ back to cell $n$ triggers.
\end{itemize}  
%
A \ac{HO} attempt can either fail or succeed. The handover failures includes \ac{HOL}, \ac{HOE} and \ac{HOW} since they all cause \acp{RLF}, while the successful handovers includes \acp{HOPP}, although they cause unnecessary handovers. Let the counts of \ac{HO} attempts, successful \acp{HO}, \acp{HOL}, \acp{HOE}, \acp{HOW}, and \acp{HOPP} for slice $s$ and boundary $(n,m)$ be denoted by $N^{\ATT}_{n,m,s}, N^{\SHO}_{n,m,s}, N^{\LHO}_{n,m,s}, N^{\EHO}_{n,m,s}, N^{\WHO}_{n,m,s}, N^{\PP}_{n,m,s}$, respectively. We propose to include them in the \ac{KPI} reports for each slice $s\in \Ss$ and each boundary $(n,m)\in\Bs$.

We define the following two per-slice {\bf \ac{HO} metrics to minimize},  {\it \ac{HFR}} and {\it \ac{PPR}}:
\begin{align}
R^{\HF}_s & = \frac{\sum_{(n,m)\in\Bs} N^{\LHO}_{n,m,s} + N^{\EHO}_{n,m,s} + N^{\WHO}_{n,m,s}}{\sum_{(n,m)\in\Bs}N^{\ATT}_{n,m,s}} \label{eqn:HFR}\\
R^{\Pp}_s & = \frac{\sum_{(n,m)\in\Bs}N^{\PP}_{n,m,s}}{\sum_{(n,m)\in\Bs}N^{\SHO}_{n,m,s}}. \label{eqn:SHOR}
\end{align}
Since $N^{\ATT}_{n,m,s} = N^{\SHO}_{n,m,s} + N^{\LHO}_{n,m,s} + N^{\EHO}_{n,m,s} + N^{\WHO}_{n,m,s}$ and $N^{\PP}_{n,m,s}\leq N^{\SHO}_{n,m,s}$, we have $R^{\HF}_s\in [0, 1]$ and $R^{\Pp}_s\in [0, 1]$, $\forall s\in\Ss$.

\subsection{Slice-Specific Service Quality Metrics}\label{subsec:ServiceMetrics}
As reflected in \eqref{eqn:Slice_HO}, the above-introduced \ac{HO} metrics are signal quality based. To observe the per-slice service performance, we define in below two {\bf service metrics to maximize}, {\it \ac{TSL}} $L^{\TS}_{k,s}$ and {\it \ac{LSL}} $L^{\LS}_{k,s}$ for each user $k$ associated to slice $s\in\Ss$:
\begin{equation}
L^{\TS}_{k,s}  = \min\left\{\frac{\eta_{k,s}}{\eta^{\ast}_s}, 1 \right\}\mbox{ and }
L^{\LS}_{k,s}  = \min\left\{\frac{d^{\ast}_s}{d_{k,s}}, 1\right\} \label{eqn:ue_SLs},
\end{equation}
where $\eta_{s, k}$ and $d_{s, k}$ are the achieved throughput and latency of user $k$ associated to slice $s$, respectively. Note that $L^{\TS}_{s,k}$ and $L^{\LS}_{s,k}$ are upper bounded by $1$. Thus, as long as the service quality is satisfied, we have $L^{\TS}_{s,k}= L^{\LS}_{s,k} = 1$.

Let the set of users associated to cell $n$ and slice $s$ be denoted by $\Ks_{n,s}$ with the cardinality $|\Ks_{n,s}| = K_{n,s}$. 
Let each cell $n$ report to the central agent the following three measures: 1) number of the users in each slice $K_{n,s}$, 2) the per-slice sum \ac{TSL} $L^{\TS}_{n,s}:=\sum_{k\in\Ks_{n,s}}L^{\TS}_{k,s}$, and 3) the per-slice sum \ac{LSL} $L^{\LS}_{n,s}:=\sum_{k\in\Ks_{n,s}}L^{\LS}_{k,s}$. The slice-specific average \ac{TSL} and \ac{LSL} for the considered area is given by
\begin{equation}
L^{\TS}_s = \frac{\sum_{n\in\Ns} L^{\TS}_{n,s}}{\sum_{n\in\Ns}  K_{n,s}} \mbox{ and } L^{\LS}_s = \frac{\sum_{n\in\Ns} L^{\LS}_{n,s}}{\sum_{n\in\Ns}  K_{n,s}}.
\label{eqn:TSL_LSL}
\end{equation}
With \eqref{eqn:ue_SLs}, we have $L^{\TS}_s\in [0, 1]$ and $L^{\LS}_s \in [0, 1]$, $\forall s\in\Ss$.

\subsection{Markov Decision Process Model}\label{subsec:MDP}
We model the multi-cell system as a \ac{MDP}, defined by the tuple $\left(\Ss, \As, P(\cdot), r(\cdot), \gamma\right)$, where $P:\Ss\times \As \times \Ss \to [0, 1]$ indicates the transition dynamics by a conditional distribution over the state space $\Ss$ and the action space $\As$, $r:\Ss\times \As\to\R$ denotes the reward function, and $\gamma\in[0,1]$ is the discount factor. We define the action, state, and reward of the \ac{MDP} as follows.

\subsubsection{Action}
Assume \ac{HOM} $O_{n,m,s}\in\Os$ and $T_{n,m,s}\in\Ts$, where $\Os$ and $\Ts$ are in practice both finite discrete sets. The $2BS$-dimensional action $\va\in\As$ is defined as
\begin{align}
\va := & [(O_{n,m,s}, T_{n,m,s}): O_{n,m,s}\in\Os, T_{n,m,s} \in\Ts, \nonumber\\
&  (n, m)\in\Bs, s\in\Ss] \in \As:=(\Os\times \Ts)^{BS}.
\label{eqn:action}
\end{align}

\subsubsection{State}
As shown in Table \ref{tab:state}, the network state $\vs \in \R^{(4N+2B)S}$ includes $4NS$ per-cell per-slice measurements and $2BS$ per-cell-pair per slice \ac{HO} event counts (with $B/2$ cell-pairs, we have $4*(B/2)*S=2BS$ per-cell-pair per-slice states). Note that we use the counts of \ac{HO} events directly instead of \ac{HFR} and \ac{PPR} defined in Section \ref{subsec:HOMetrics} because of two reasons: 1) sometimes the number of HO attempts is zero and using raw counts avoids zero denominators, and 2) raw counts may provide more information reflecting too-early and too-late \ac{HO} decisions.  Also, we reduce the dimension of state by adding the \ac{HO} event counts of the both directions of the cell boundaries. For the same reason we do not include the counts of \ac{HOE} and \ac{HOW} into the state, since we expect that the count of \acp{HOPP} indicates the too-early \ac{HO} decisions.

\subsubsection{Reward}
The reward function is a utility function based on the \ac{HO} metrics $R^{\HF}_s$ and $R^{\Pp}_s$ defined in \eqref{eqn:HFR} and \eqref{eqn:SHOR} respectively, and the service metrics $L^{\TS}_s$ and $L^{\LS}_s$, for $s\in\Ss$ defined in \eqref{eqn:TSL_LSL}.  For example, one option is given by:
\begin{equation}
r = \sum_{s\in\Ss, X\in\{\operator{T}, \operator{D}\}} w_s^{(X)} L_s^{(X)} - \sum_{s\in\Ss, X\in\{\operator{HF}, \operator{PP}\}} w_s^{(X)} R_s^{(X)}.
\label{eqn:reward}
\end{equation}
where $w_s^{(X)}$ for $s\in\Ss$, $X\in\{\operator{T}, \operator{D}, \operator{HF}, \operator{PP}\}$ are the weight factors. Note that $L_s^{\LS}$ is the latency satisfaction level that is {\bf inversely} proportional to latency -- a larger value indicates higher satisfaction by meeting the latency constraints \eqref{eqn:ue_SLs}. The designed reward function incentives to improve the service quality satisfaction level, as well as to minimize the handover failures and unnecessary handovers. Because all the metrics included in reward are within $[0, 1]$, we have the same scale for the metrics, and the reward is bounded. 
 %
\begin{center}
\begin{table}[tbp]
 \caption{Measures included in the state}\label{tab:state}
\begin{tabular}{lllll}
\cline{1-2}
\multicolumn{1}{|l|}{Per-cell per-slice states} & \multicolumn{1}{l|}{Per-cell-pair per-slice states} &  &  &  \\ \cline{1-2}
\multicolumn{1}{|l|}
{\begin{tabular}[c]{@{}l@{}}
$\forall n\in\Ns, s\in\Ss:$ \\
$\cdot$ Load $l_{n,s}$\\ 
$\cdot$ Number of users $K_{n,s}$\\ 
$\cdot$ TSL $L_{n,s}^{\TS}$\\ 
$\cdot$ LSL $L_{n,s}^{\LS}$ \end{tabular}} & 
\multicolumn{1}{l|}
{\begin{tabular}[c]{@{}l@{}}
$\forall (n,m)\in \Bs', s\in \Ss:$\\
$\cdot$ HO attempts $N^{\ATT}_{n,m,s} + N^{\ATT}_{m,n,s}$ \\ 
$\cdot$ Successful HOs $N^{\SHO}_{n,m,s} + N^{\SHO}_{m,n,s}$\\ 
$\cdot$ \acp{HOL} $N^{\LHO}_{n,m,s} + N^{\LHO}_{m,n,s}$\\ 
$\cdot$ \acp{HOPP} $N^{\PP}_{n,m,s} + N^{\PP}_{m,n,s}$\end{tabular}}
 &  &  &   \\ \cline{1-2}   &  &  &  & \\   &  &  & &
\end{tabular}
\end{table}
\end{center}
\section{Problem Statement}\label{sec:Problem}
As many classical reinforcement learning problems, we want to find a policy $\pi:\Ss\to\As$ which decides the \ac{HO} parameters $\va_t$ based on the network state $\vs_t$, to maximize the expectation of the cumulative discounted reward of a finite horizon $H$:
\begin{equation}
\maximize_{\pi} \  \Expt_{\pi}\left[\sum_{t=0}^H\gamma^t r(\vs_t, \va_t)\right], \mbox{ s.t. } \va\in \As.
\label{eqn:problem}
\end{equation}
However, we face the following two challenges. The first one is the complexity caused by the enormous discrete action space. In 3GPP, it is defined $\Os := \{-24, -23, \ldots, 23, 24\}$ (in dB) and $\Ts := \{0, 40, 64, 80, 100, 128, 160, 256, 320, 480, 512, 640, 1024,\\ 1280, 2560, 5120\}$ (in ms) \cite{TS38331}, which means $|\Os| = 49$ and $|\Ts| = 16$. Even for a neighboring area of $9$ cells, assuming that each cell only has $2$ neighboring cells, we have $36$ directional cell boundaries. With only $2$ slices per cell, the size of the action space yields $|\As|=(49*16)^{72}> 1e208$. This is intractable with the value-approximation methods for discrete action space such as \ac{DQN}.  

The second challenge is raised by the high cost of online interaction. Reinforcement learning, especially with large state and action spaces, usually requires sufficient exploration to converge. However, in real-world network systems, executing actions leading to unknown states may trigger risky network states. On the other hand, it is possible to collect an off-policy sample set $\left\{(\vs_t, \va_t, \vs_{t+1}, r_t): t\in\N_0\right\}$ in the safe operating action space. Thus, our problem is to learn from a biased offline sample set with a minimal number of online interactions.

\section{Proposed Approach}\label{sec:Solution}
To deal with the large discrete action space, we propose a policy built upon the actor-critic algorithms with a projection policy that maps the continuous proto-action to its nearest neighbors in the discrete action space $\As$. We  describe \ac{TD3} \cite{fujimoto2018addressing} as the applied actor-critic algorithm in Section \ref{subsec:Actor-critic}. Then, we propose the action projection policy in Section \ref{subsec:DiscAct}. To learn from the biased data, we propose in Section \ref{subsec:TFlearning} a transfer learning approach for \ac{DRL} that comprises two steps: 1) regularized offline \ac{DRL}, and 2) online \ac{DRL} model fine-tuning.  
 
\subsection{Actor-Critic Algorithm}\label{subsec:Actor-critic}
The objective function in \eqref{eqn:problem} can be seen as a state-action value function $Q^{\pi}(\vs, \va)=\Expt^{\pi}\left[\sum_{t=0}^H\gamma^t r_t|\vs_0=\vs, \va_0=\va\right]$. The value function $Q^{\pi}$ can be expressed in a recursive manner using the Bellman equation \cite{bertsekas2012dynamic}:
\begin{equation}
Q^{\pi}(\vs, \va) = r(\vs, \va) + \gamma\sum_{\vs'|\vs, \va}Q^{\pi}(\vs', \pi(\vs')).
\label{eqn:Bellman}
\end{equation} 
For value-based \ac{DRL} in discrete space, $Q^{\pi}(\vs, \va)$ is approximated by a neural network which takes both state and action as input, and output $|\As|$ evaluations of $Q$ values corresponding to every possible action. Then, it chooses the action with the best estimation of the Q value to execute. Unfortunately, because its complexity grows linearly with $|\As|$, it is intractable when the discrete action space is huge.  

We consider next relaxing the action space to the continuous space, which enables us to apply the actor-critic method in the continuous action space. More details of action projection will be provided in Section \ref{subsec:DiscAct}.

The actor-critic method solves the optimization problem by using a critic $Q_{\theta}\left(\vs_t, \va_t\right)$ to approximate the value function, i.e., $Q_{\theta}(\vs_t, \va_t) \approx Q^\pi(\vs_t, \va_t)$ and an actor $\pi_{\omega}(\vs_t)$ to update the policy $\pi$ at every step in the direction suggested by critic. The critic and actor both modeled with neural networks are characterized by parameters $\theta$ and $\omega$, respectively. In this way, unlike \ac{DQN} having $|\As|=(|\Os||\Ts|)^{BS}$ outputs for our problem formulation (it estimates Q values corresponding to all possible actions), the actor has an output dimension of $2BS$ because it directly predicts the action $\va\in\R^{2BS}$.

We follow a particular actor-critic algorithm, namely, the \ac{TD3} \cite{fujimoto2018addressing}, to train the agents, which has proven effectiveness when dealing with high dimensional and continuous state space. As an extension of \ac{DDPG} \cite{silver2014deterministic}, \ac{TD3} overcomes the \ac{DDPG}’s problem of overestimating Q values by introducing double critic structure for both current networks $Q_{\theta_1}, Q_{\theta_2}$ and target networks $Q_{\theta'_1}, Q_{\theta'_2}$. The minimum of the two Q values is used to represent the approximated Q value of the next state. Moreover, \ac{TD3} uses smoothing for target policy $\pi_{\omega'}$, i.e., adding noise to the target action to make it harder for the policy to exploit Q-function errors by smoothing out Q along changes in action. The target actions are computed by
\begin{equation}
    \label{eqn:target_action}
    \va'(\vs_{t+1}) = clip\left(\pi_{\omega'}\left(\vs_{t+1}\right) + clip(\epsilon, -c, c), \va_{\operator{L}}, \va_{\operator{H}}\right),
\end{equation}
where the added noise $\epsilon\sim \N(0, \sigma)$ is clipped within $[-c, c]$ to keep the target close to the original action, and $\va_{\operator{L}}, \va_{\operator{H}}$ are the lower and upper bounds of the action, respectively. 

The critics' parameters $\theta_i, i=1,2$ are updated with temporal difference (TD) learning, given by:
\begin{equation}
\label{eqn:update-critic}
\begin{aligned}
	L\left(\theta_i\right) & = \Expt\left[\left(y_t - Q_{\theta_i}\left(\vs_t, \va_t\right)\right)^2\right], i =1,2\\
\mbox{where }	y_t & = r_t + \gamma \min_{i=1,2} Q_{\theta'_i}\left(\vs_{t+1}, \va'(\vs_{t+1})\right).
\end{aligned}
\end{equation}

The actor is updated with the policy gradient on the expected accumulated regularized reward, give by
\begin{equation}
\Expt\left[\nabla_{\va}Q_{\theta_1}(\vs,\va)|_{\vs=\vs_t, \va=\pi_{\omega}(\vs_t)}\nabla_{\omega}\pi_{\omega}\left(\vs\right)|_{\vs=\vs_t}\right].\label{eqn:actorgradient}
\end{equation}

\subsection{Dealing with Large Discrete Action Space}\label{subsec:DiscAct} 
In this part, we propose the method to relax the discrete action space to the continuous space, as well as to map the proto-action back to the discrete action by approximating its nearest neighbors in $\As$.
 
Motivated by \cite{dulac2015deep}, given the proto-action $\tilde{\va}$ as the output of actor $\pi:\Ss\to\R^{2BS}$, it is very likely that $\tilde{\va}\notin\As$. We need to design a policy to map $\tilde{\va}$ to an action in the desired discrete space $\hat{\va}\in\As$. 
The mapping policy is similar to the {\it Wolpertinger Policy} in \cite{dulac2015deep}, summarized in Algorithm \ref{alg:Wolpertinger}. We first find with policy $G_k$ (Algorithm \ref{alg:neighborselection}) the $k$ nearest neighbors of $\tilde{\va}$ in $\As$, denotes by $\As_k(\tilde{\va}):=\left\{\va^{(1)}, \ldots, \va^{(k)}\right\}$, then select the one with the best $Q$-estimates (estimated by the critic function) as the action applied to the environment. The mapped $\hat{\va}\in\As$ is given by:
\begin{equation}
\hat{\va} = \argmax_{\va\in G_k\circ \pi_{\omega}(\vs)} Q_{\theta_1}\left(\vs, \va\right).
\label{eqn:mapping_Q}
\end{equation}
\begin{algorithm}
\caption{Action Projection Policy}
\label{alg:Wolpertinger}
\begin{algorithmic}[1]
\State Get state $\vs$
\State $\tilde{\va}=\pi_{\omega}(\vs)$  \Comment{Get proto-action from actor}
\State $\As_k(\tilde{\va}) = G_k(\tilde{\va})$ \Comment{Get $k$ nearest neighbors of $\tilde{\va}$ }
\State $\hat{\va}=\argmax_{\va\in\As_k(\tilde{\va})} Q_{\theta_1}\left(\vs, \va\right)$ 
\State Apply $\hat{\va}$ to environment; receive $r, \vs'$.
\end{algorithmic}
\end{algorithm}
As for the strategies of selecting $k$ nearest neighbors $G_k:\R^{2BS}\to\As^k$, in \cite{dulac2015deep} the authors use approximate nearest neighbor methods based on K-Means tree structure \cite{muja2014scalable} allowing for logarithmic-time search complexity relative to the size of action space $O\left(ZBS (\log|\As|/\log k)\right)$, where $Z$ is the maximum number of points to examine (a hyperparameter of the algorithm). However, except for the search complexity, the tree construction complexity is also substantially high and requires extra large memory space. Considering that our action space is already well structured (containing gridded points in high-dimensional space), we propose a coarse, heuristic, but more efficient way to approximate the $k$ neighbors with the search complexity $O(kBS(|\Os|+|\Ts|))$, described in Algorithm \ref{alg:neighborselection}. Let the $j$-th dimension of $\tilde{\va}$ be denoted by $\tilde{a}_j$. The intuition is to choose the nearest values in $\Os$ or $\Ts$ along each dimension with the probability depending on the reciprocal of the distance from $\tilde{a}_j, \forall j$. The proposed heuristics can quickly generate $k$ neighbors of $\tilde{\va}$ in $\As$. 
\begin{algorithm}
\caption{Approximate $k$ neighbor selection}
\label{alg:neighborselection}
\begin{algorithmic}[1]
\State $\As_k(\tilde{\va})\leftarrow\emptyset$
\For{the $i$-th neighbor to be generated, $i\in\{1, \ldots, k\}$}
	\For{$j\in\{1, \ldots, 2BS\}$}
	\State $\Omega \leftarrow \Os$ if the dimension is \ac{HOM}, o/w $\Omega \leftarrow \Ts$
		\If{$\tilde{a}_j>=\max(\Omega)$ or $\tilde{a}_j<=\min(\Omega)$}
		\State $a^{(i)}_j \leftarrow \argmin_{x\in\Omega} |x-\tilde{a}_j|$
		\Else
		\State $a_j^{+} \leftarrow \argmin_{x\in\Omega, x\geq \tilde{a}_j} |x-\tilde{a}_j|$
		\State $\nu_j^{+} \leftarrow 1/(a_j^{+} - \tilde{a}_j)$
		\State $a_i^{-} \leftarrow \argmin_{x\in\Omega, x\leq \tilde{a}_j} |x-\tilde{a}_j|$
		\State $\nu_j^{-} \leftarrow  1/(\tilde{a}_j - a_j^{-})$
		\State $p_j^{+} \leftarrow \nu_j^{+}/(\nu_j^{+} + \nu_j^{-})$; $p_j^{-} \leftarrow 1-p_j^{+}$
		\State $\Pr\left\{a^{(k)}_j \leftarrow a_j^{+}\right\}=p_j^{+}$  and 
		\State $\Pr\left\{a^{(k)}_j \leftarrow a_j^{-}\right\}=p_j^{-}$
		\EndIf
	\EndFor 
	 \State $\va^{(i)} \leftarrow \left[a_1^{(i)}, \ldots, a_{2BS}^{(i)}\right]$
	\State $\As_k(\tilde{\va})\leftarrow \As_k(\tilde{\va})\cup \{\va^{(i)}\}$
\EndFor
\end{algorithmic}
\end{algorithm}
\subsection{Transfer Learning in Deep Reinforcement Learning}\label{subsec:TFlearning}
The proposed transfer learning approach comprises of two steps: 1) regularized offline actor-critic training, and 2) online fine-tuning with mixed replay buffer sampling.

\subsubsection{Regularized Offline Deep Reinforcement Learning}\label{subsec:ReOffDRL}
Off-policy batch reinforcement learning \cite{gao2020batch} performs the task of learning from a collected dataset $\Ds:=\left\{\left(\vs^{(i)}, \va^{(i)}, \vs'^{(i)}, r^{(i)}\right): i = 1, \ldots, D\right\}$ without further interactions with the environment. 
In the Telecom industries, however, the available data usually contains network configurations in a safe operating space. Such biased dataset causes the algorithms to fail because of a fundamental problem of {\it extrapolation error}, a phenomenon in which the model erroneously over-estimates the Q values of the unseen state-action pairs, and outputs the actions with over-estimated unrealistic rewards (see Fig. \ref{fig:extrapolation}). 
To overcome the challenge, we propose to add a regularization term to the reward function, such that the probability of the state-action pair is also maximized. Note that this motivation is opposite to some of the online algorithms that encourage the exploration by maximizing the expected entropy of the policy. The rationale is that, for the offline optimization without interaction with the environment, exploring outside of training data distribution leads to unrealistic over-estimated values.      

We replace the reward $r$ in \eqref{eqn:problem} and \eqref{eqn:Bellman} by
\begin{equation}
r'(\vs_t, \va_t) = r(\vs_t, \va_t) + \alpha\log p(\vs_t, \va_t),
\label{eqn:mofdifiedRew}
\end{equation}  
where $p(\vs_t, \va_t)$ is the probability density function that $(\vs_t, \va_t)$ appears in the training dataset, and $\alpha$ is a weight factor.
%
The remaining problem is how to compute the log-density estimation  $\log p(\vs_t, \va_t)$. Density estimation for complex high-dimensional data is a challenging fundamental problem in statistical learning. Assuming that we observe a random vector $\vx\in\R^d$ and want to use a parametrized density model to approximate the log-density function $\log p_{\vx}(\cdot)$. An efficient method is the energy-based models, i.e., to define an energy function $\Es_{\vx}(\cdot)$ which is essentially an unnormalized log-density function for the given data, and define the score function $\phi(\vx) = \nabla_{\vx}\log p_{\vx}(\vx) = -\nabla_{\vx}\Es_{\vx}(\vx)$. The energy function can be then estimated using {\it score matching} \cite{hyvarinen2005estimation}. In this work, we apply a promising denoising score matching-based solution, \ac{DEEN}, which uses a neural network to model the energy $\Es_{\psi}\left(\vx\right)$ characterized by parameters $\psi$. 
Due to the limited space, we do not give full details but refer the interested readers to \cite{saremi2018deep}.  The regularized reward in \eqref{eqn:mofdifiedRew} is then given by
\begin{equation}
r'\left(\vs_t, \va_t|\psi\right) = r(\vs_t, \va_t) - \alpha \Es_{\psi}\left(\vs_t, \va_t\right)
\label{eqn:rwd_Energy}
\end{equation} 

It is also worth noting that the size of the offline dataset $\Ds$ is limited, and  the collected operating actions are discrete. To fully exploit them to train the offline \ac{DRL} whose outputs of actor are continuous, we can use the inverse policy of the neighbor selection strategy in Algorithm \ref{alg:neighborselection} for the data augmentation of the training set. Similar to generating the discrete neighbors of a proto-action, we generate the $k$ neighboring continuous actions $\tilde{\As}_k(\va):=\{\tilde{\va}^{(i)}: i = 1, \ldots, k\}$ near the discrete action in the dataset $\va\in\As$. 
In this way, from each sample $\left(\vs, \va, \vs', r\right)\in\Ds$, we augment $k$ samples with actions in the continuous space $\{\left(\vs, \tilde{\va}, \vs', r\right): \forall \tilde{\va}\in \tilde{\As}_k(\va)\}$, and the training dataset is extend to $\overline{\Ds}$ with size $kD$. 

\subsubsection{The Online Fine-tuning of the Pretrained Models}\label{subsec:Finetuning}
We expect that the offline pretrained actor-critic agent provides us a good initialization of the online operation. However, to adapt to the real-time system that may have a distribution drift from the training data, we allow the model to fine-tune itself with the real-time interaction with the environment. To this end, we propose to build two replay buffers: $B^{\off}$ stores offline training data, while $B^{\on}$ stores the updated new samples. For each training update of the online agent, the mini-batch can be sampled from either $B^{\off}$ or $B^{\on}$. During the training, we gradually decrease the probability of sampling from $B^{\off}$, while increase the probability of sampling from $B^{\on}$. Meanwhile, along with the actor and critic networks, we also periodically update the energy estimation network $\Es_{\psi}\left(\vx\right)$ based on the updated offline and online datasets. 
The TD3-based SAMRO algorithm is then provided in Algorithm \ref{algo:onlineRL}.  Note that \ac{TD3} builds two critic networks and takes the minimum value between them to limit the over-estimation of the Q value. 
\begin{algorithm}
\caption{TD3-based SAMRO with knowledge transfer}
\label{algo:onlineRL}
\begin{algorithmic}[1]
\Initialize{Critic networks $Q_{\theta_i}$, $i=1,2$ \\
Actor network $\pi_{\omega}$\\
Target critic networks $Q_{\theta'_i}$, $i=1,2$\\
Target actor network $\pi_{\omega'}$\\
Energy estimator network $\Es_{\psi}$\\
Replay buffers $B^{\off}\neq\emptyset$ and $B^{\on}\leftarrow \emptyset$\\
Buffer prioritization factor $\beta$
} 
\For{$t=1,\ldots, T$}
\State Output action with exploration noise $\tilde{\va}\leftarrow \pi_{\omega}(\vs) + \epsilon$
\State Select operating action $\hat{\va}$ using Algorithm \ref{alg:Wolpertinger}
\State Observe $r$ and new state $\vs'$
\State Store transition tuple $(\vs, \tilde{\va}, \vs', r)$ in $B^{\on}$
\State Sample mini-batch of $M$ transitions from $B^{\on}$ with 
\State probability $\beta$ while from $B^{\off}$ with probability $1-\beta$
\State $\tilde{\va}'\leftarrow \pi_{\omega'}(\vs')$
\State Update $r'$ with \eqref{eqn:rwd_Energy}
\State $y\leftarrow r' + \gamma \min_{i=1,2}Q_{\theta_i'}\left(\vs', \tilde{\va}'\right)$
\State Update critics' weights $\theta_i$, $i=1,2$ by minimizing
\State the sampled loss of \eqref{eqn:update-critic}
\If{$t\mod H^{\Act}$}
\State Update actor weights $\omega$ using the sampled \State gradient of \eqref{eqn:actorgradient} and update the target networks' 
\State weights $\theta'_i$, $i=1,2$ and $\omega'$
\EndIf
\If{$t\mod H^{\Eg}$}
\State Update energy estimator network weights $\psi$ 
\State Update $\beta$   
\EndIf
\EndFor
\end{algorithmic}
\end{algorithm}
\section{Numerical Results}\label{sec:Numerical}
We test the SAMRO algorithm proposed in Section \ref{sec:Solution} with a realistic system-level \ac{SON} simulator, as an extension with network slicing of the emulator in \cite{network2009introducing}.

\subsubsection{Simulator Scenario}\label{subsec:simscenario}
 \begin{figure}[t]
    \centering
    \includegraphics[width=0.48\textwidth]{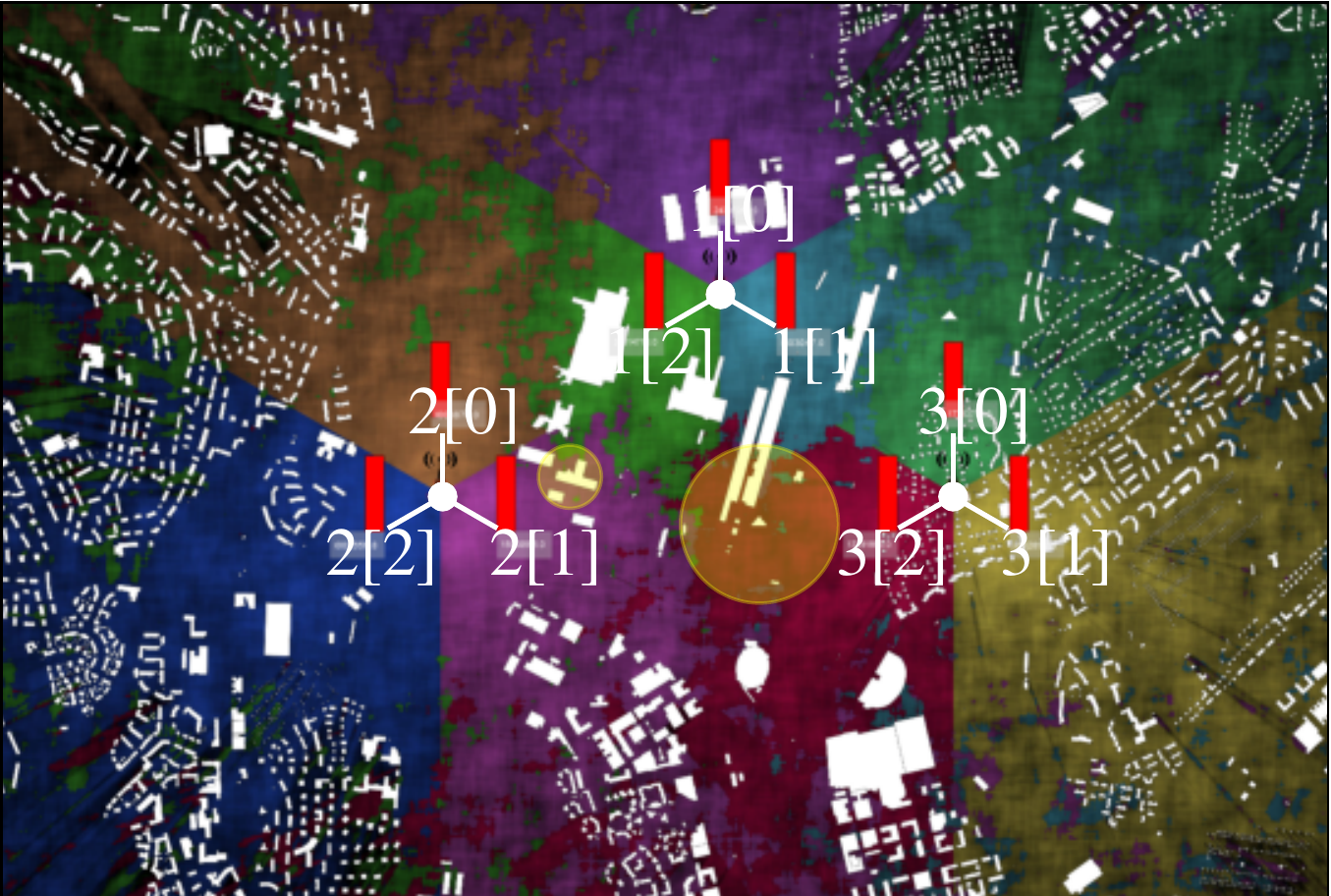}
    \caption{The $9$-cell simulation environment. Red bar indicates the cell throughput at the snapshot. White areas are the buildings. Yellow circles define the mobility areas of the hotspots.}
    \label{fig:geo}
\end{figure}
\begin{figure}[t]
    \centering
    \includegraphics[width=0.48\textwidth]{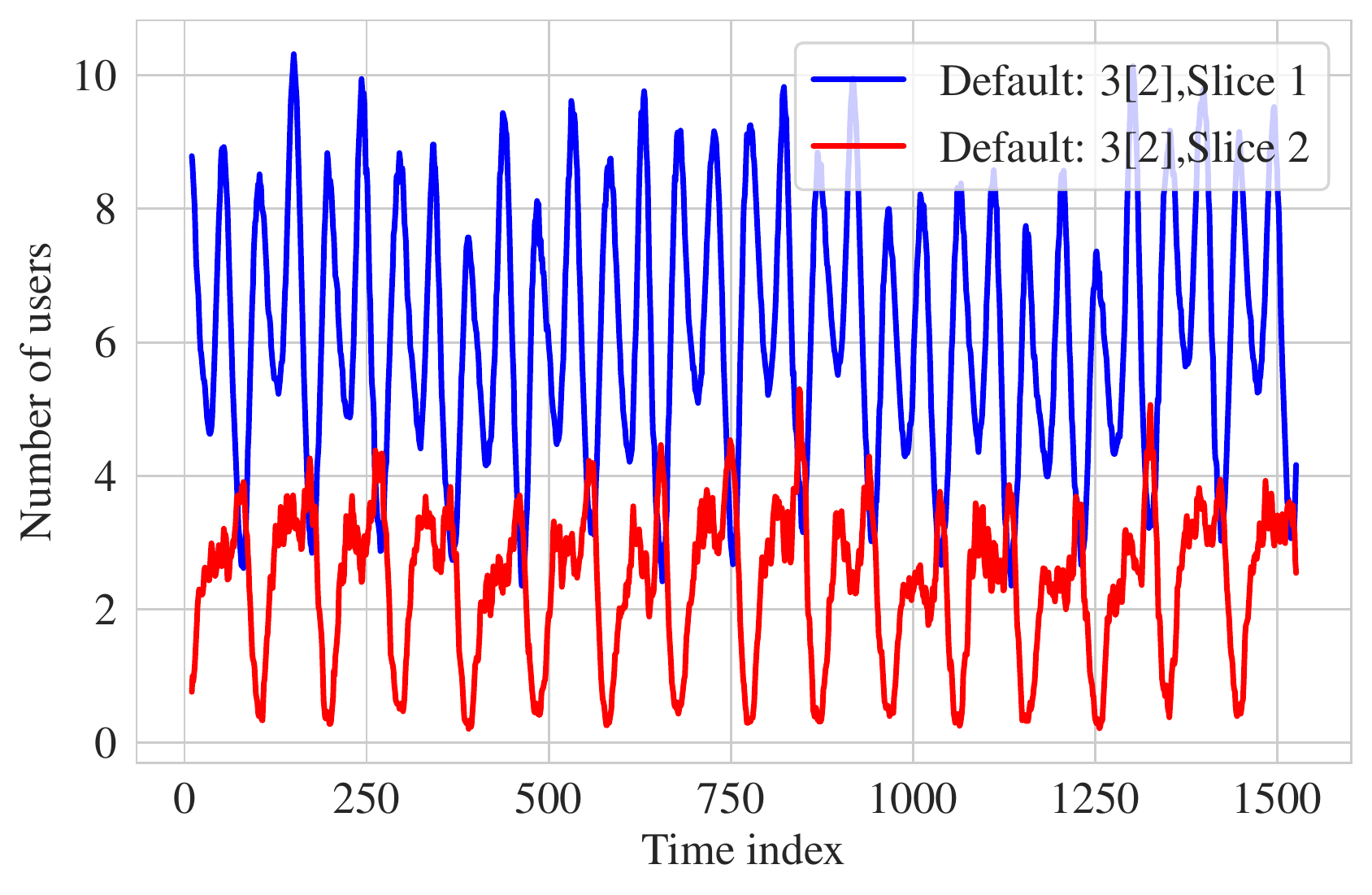}
    \caption{Number of connected users in Cell \lq\lq 3[2]\rq\rq \ with hotspot when \ac{HO} parameters are set to default.}
    \label{fig:traffic}
\end{figure}

We consider an environment with $3$ three-sector sites, i.e., $9$ cells, operating on the frequency band $2.4$ GHz, as shown in Fig. \ref{fig:geo}. The radio propagation model is Winner+ \cite{meinila2010d5} supported with ray tracing.  In Table \ref{tab:UE_group}, we define four groups of users associated to two slices: \lq\lq Slice 1\rq\rq \ supporting video traffic and  \lq\lq Slice 2\rq\rq \ supporting HTTP traffic. The users in groups $1$ and $2$ move uniformly randomly within the playground and those in hotspots move within the circled areas as shown in Fig. \ref{fig:geo}. Due to the limitation of the scheduler in simulator, we assume all slices have the same latency requirement of $1$ ms but different throughput requirements. we also generate a realistic traffic pattern to each of the user group as shown in Fig. \ref{fig:traffic}. 
The agent interacts with the simulator with a time granularity of $900$ simulation slots, which reflects $15$-min of the real-life time. Namely, a one-day operation (in simulator) allows $96$ interactions with the simulated environment. 
%
\begin{table}[tbp]
\caption{User Groups}
\label{tab:UE_group}
\begin{tabular}{|l|l|l|l|l|}
\hline
          & Group size & Traffic Type & Expected Rate & Speed  \\ \hline
Group $1$   & $25$         & Slice $1$ (Video)          & $5$ Mbit/s      & 6 km/h \\ \hline
Group $2$   & $25$         & Slice $2$ (HTTP)         & $3$ Mbit/s      & 3 km/h \\ \hline
Hotspot $1$ & $8$          & Slice $1$ (Video)        & $5$ Mbit/s      & 3 km/h \\ \hline
Hotspot $2$ & $8$          & Slice $1$ (Video)        & $5$ Mbit/s      & 3 km/h \\ \hline
\end{tabular}
\end{table}
\begin{figure}[t]
    \centering
    \includegraphics[width=0.48\textwidth]{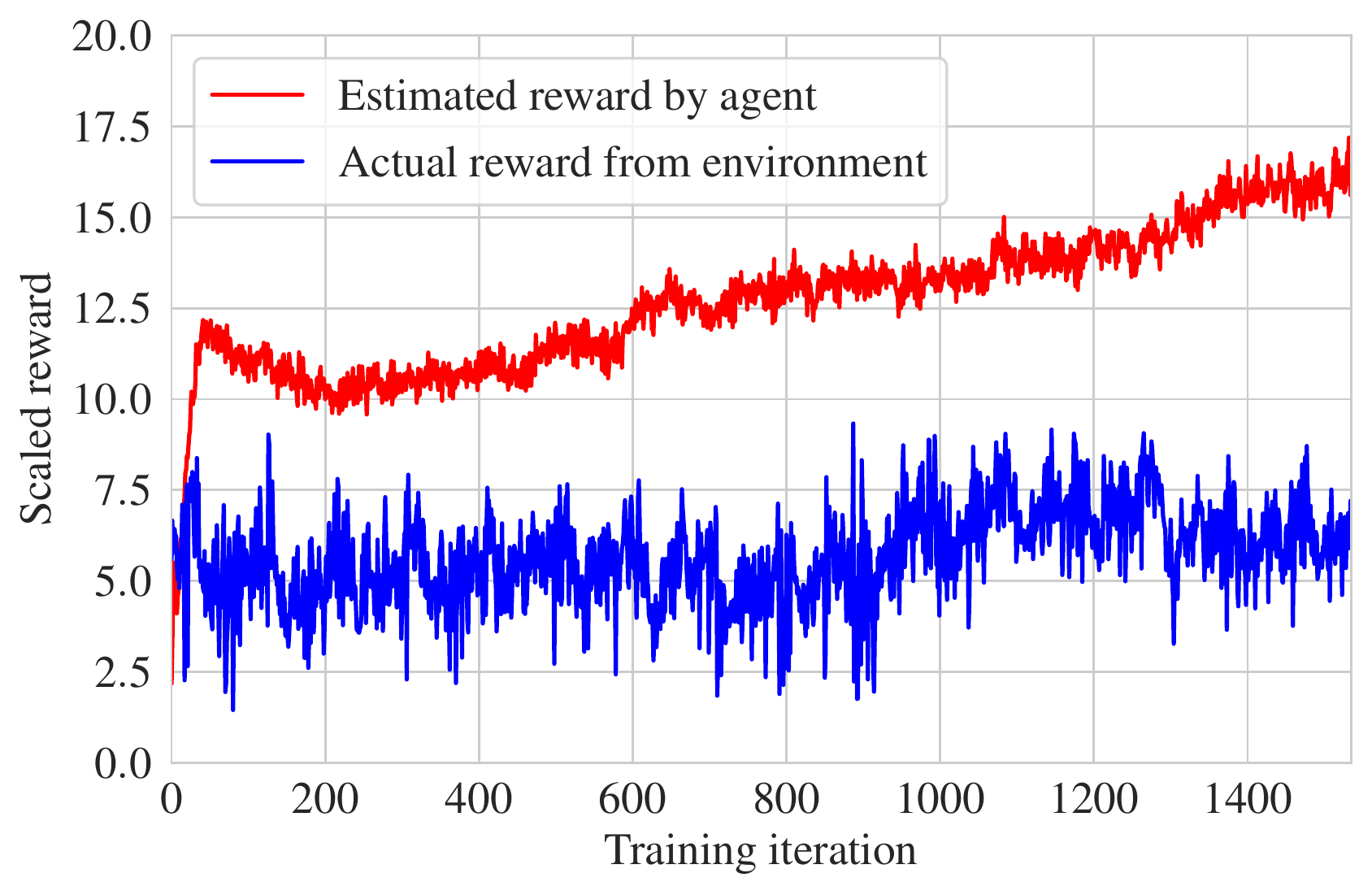}
    \caption{Phenomenon of extrapolation error.}
    \label{fig:extrapolation}
\end{figure}

\subsubsection{Offline Regularized Deep Reinforcement Learning}\label{subsec:off-policy}
Following the state and action defined in Section \ref{subsec:MDP}, we define $N=9$, $S=2$, and $B=34$. Thus, the state $\vs\in \R^{208}$ and the proto-action $\tilde{\va}\in\R^{136}$. We define $\Os:=\{-5,-4, \ldots, 5\}$ in dB and $\Ts:=\{40, 64, 80, 100, 128, 160, 256, 320, 480, 512,$ $640, 1024, 1280, 2560, 5120\}$ in ms. Thus, we also define the upper and lower bounds for each dimension of the proto-action. We assume that in practice the default value of \ac{HOM} is $0$ dB and \ac{TTT} is $512$ ms. 
We have collected $20,000$ offline biased samples as follows. Firstly, we generated independent normal distributed r.v.s for each dimension of $\tilde{\va}$ with the mean as the default value, and the variance of $3$ dB for \ac{HOM} and $300$ ms for \ac{TTT}, respectively. Then, we find the nearest value in $\Os$ or $\Ts$ and send the discrete action back to the simulator. The reward weights $w_s^{\TS}, w_s^{\LS}, w_s^{\HF}, w_s^{\Pp}$ are set as $1,1,1,0.3$, respectively. It is worth noting that sometimes scaling the reward values help improve the convergence. We scale the reward with a factor of $5$. With the defined weights, after the scaling the range of the reward is $[-6.5, 10]$. We choose small discount $\gamma=0.1$ because optimizing the network configuration for every $15$-min mainly influences the instantaneous reward. 

As for the actor-critic networks, we use \ac{MLP} for the actor and critic networks, both with $3$ hidden layers, and numbers of neurons for each layer are $(128, 64, 32)$ and $(64,16,4)$, respectively. Learning rates of the actor and critic are $0.001$ and $0.002$, respectively. The batch size is $64$, and the optimizer is Adam. For the \ac{DEEN}, we also use $3$ layers with the sizes $(256, 64, 32)$. The batch size is $32$ and the noise scale is $0.1$. The update periods of actor and \ac{DEEN} are $H^{\Act}=3$ and $H^{\Eg} = 100$ respectively.

Fig. \ref{fig:extrapolation} illustrates a typical extrapolation error when we train an offline \ac{DRL} without regularization in Fig. \ref{fig:extrapolation}. To verify this, we set the discount factor $\gamma=0$ such that the critic simply predicts the instantaneous reward and the policy should output the action with the best reward. However, after $30$ mini-batches, the agent starts over-estimating the reward for unseen actions (even exceed the upper-bound). 
%

\begin{figure}[t]
    \centering
    \includegraphics[width=0.48\textwidth]{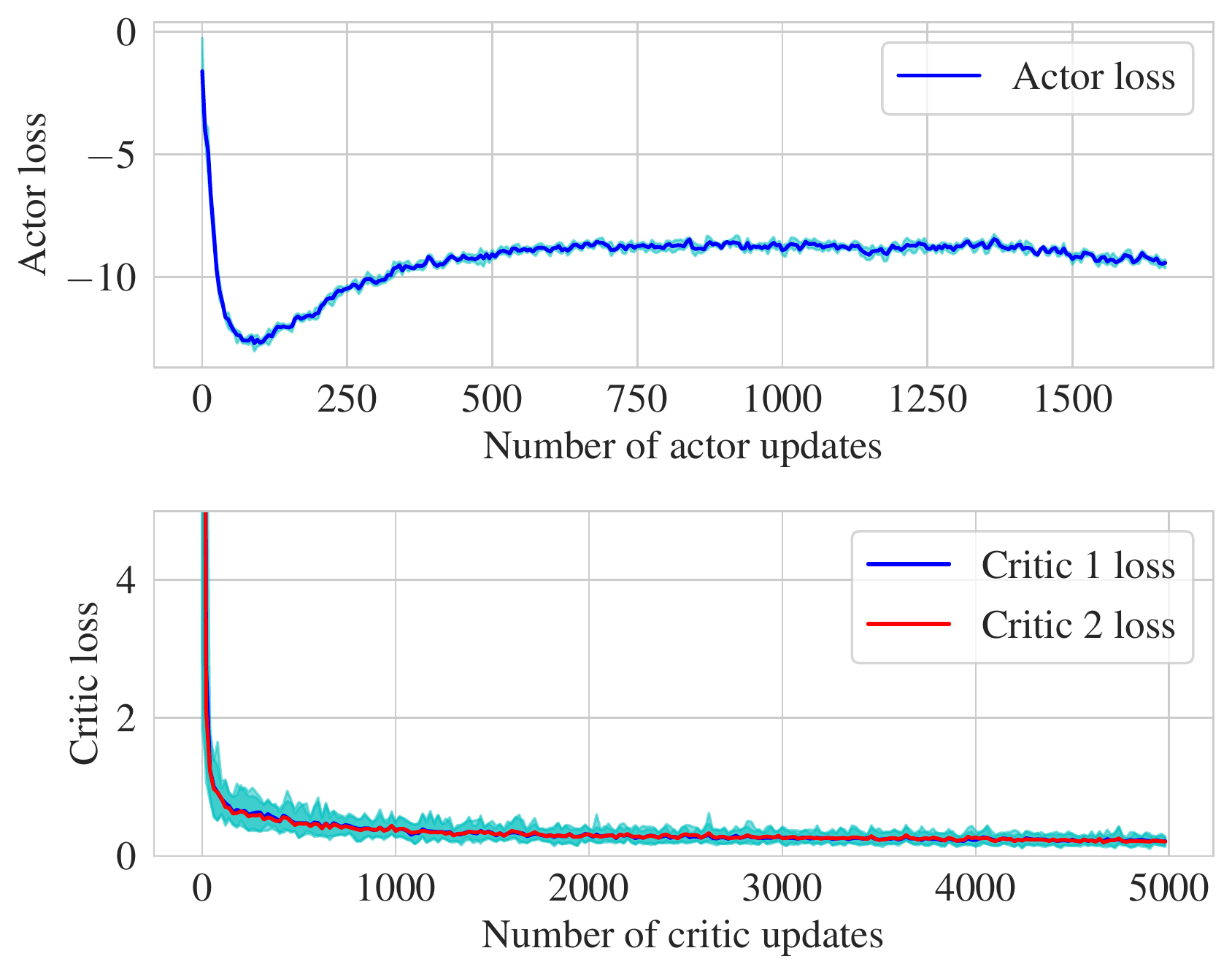}
    \caption{Training loss of actor and critic networks.}
    \label{fig:off_training}
\end{figure}

Fig. \ref{fig:off_training} shows the training loss functions of the actor and critics with the regularization. The regularized term help the negative actor loss (reflecting the Q estimate of the chosen next action) to be well bounded within the scaled reward region and prevent from outputting an action outside of the training data distribution.

\subsubsection{TD3-Based \ac{SAMRO} with Knowledge Transfer}\label{subsec:onlinetraining}
To show the advantage introduced by slice-specific \ac{HO} parameters and \acp{KPI}, we compare our \ac{SAMRO} approach with two baselines: 1) \ac{MRO} optimized with the same \ac{DRL} with transfer learning approach, but without slice-awareness, and 2) default setting of \ac{HOM} to be $0$ dB and \ac{TTT} to be $512$ ms. 

\begin{figure}[t]
    \centering
    \includegraphics[width=0.48\textwidth]{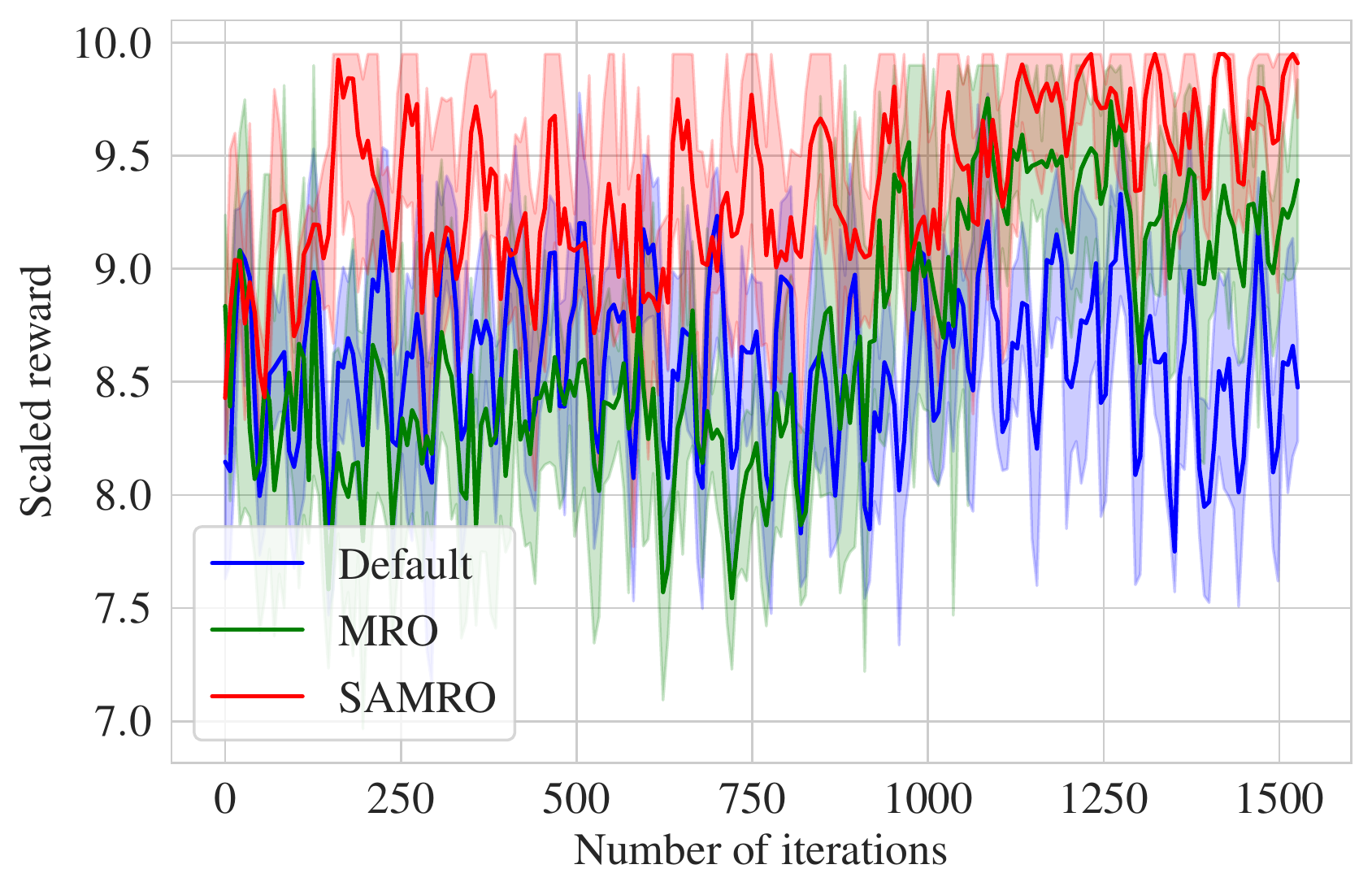}
    \caption{Online training against baselines.}
    \label{fig:base_compare}
\end{figure}

Fig. \ref{fig:base_compare} shows that our approach helps both \ac{SAMRO} and \ac{MRO} converge faster and outperform the default setting with both higher start and higher asymptote, when assuming only two weeks of online training ($1344$ training interactions) and two days of testing ($192$ testing interactions). 

\begin{figure*}[t]
    \centering
    \includegraphics[width=0.98\textwidth]{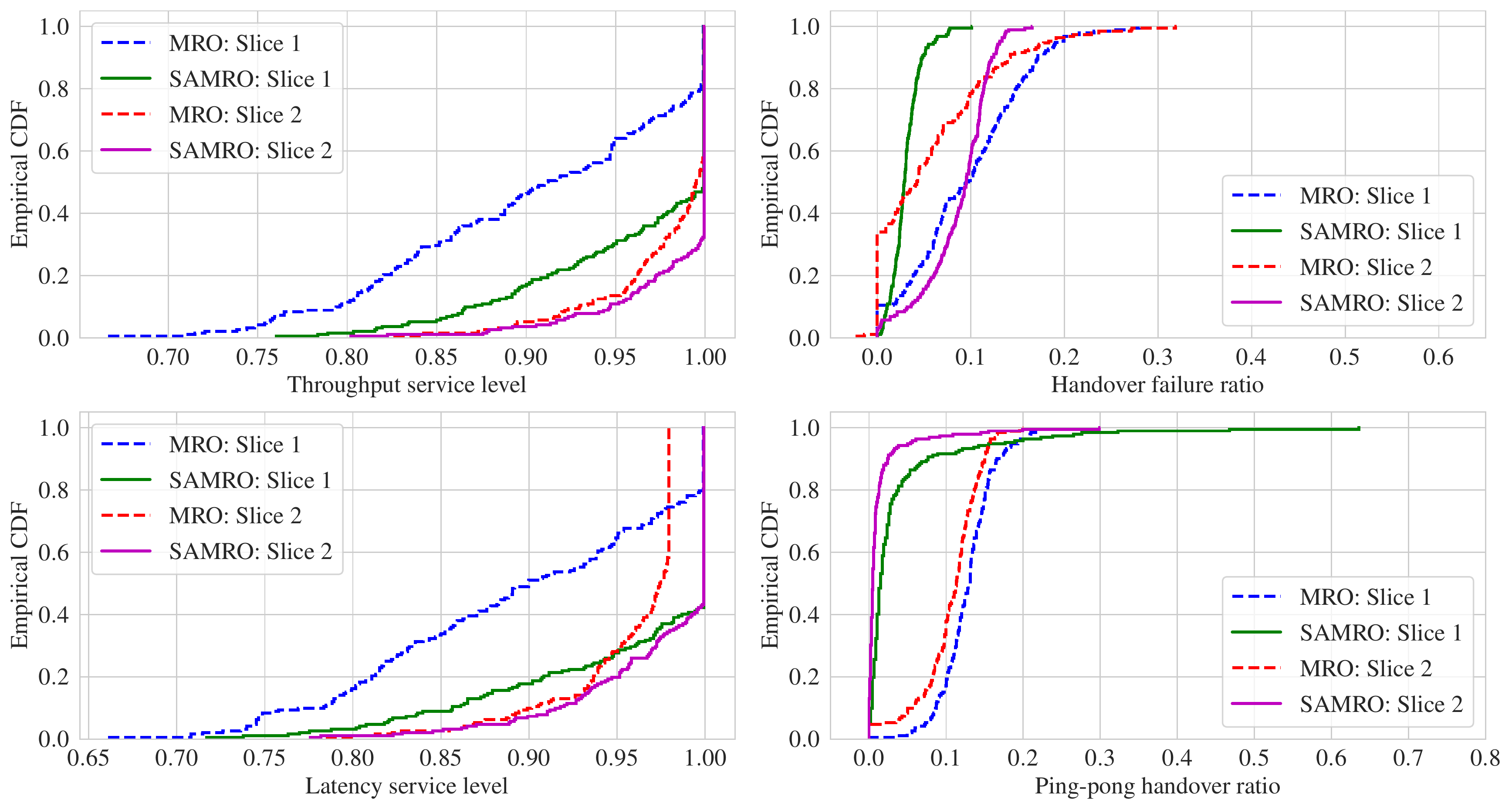}
    \caption{Comparison of the service and \ac{HO} quality.}
    \label{fig:HO_service_compare}
\end{figure*}

Fig. \ref{fig:HO_service_compare} illustrates the empirical \acp{CDF} of {\it\ac{HO} metrics \ac{HO} failure ratio} and {\it ping-long \ac{HO} ratio} and the slice service metrics {\it throughput service level} and {\it latency service level}, respectively. We observe that \ac{SAMRO} further outperforms \ac{MRO} because it not only improves the \ac{HO} performance, but also reduces the violation of the throughput and latency constraints of each slice. For example, \ac{SAMRO} guarantees that more than $55\%$ and $70\%$ of the services satisfy the throughput requirements of Slice $1$ and $2$, respectively, while \ac{MRO} satisfies only less than $20\%$ and $40\%$, respectively. Moreover, \ac{SAMRO} achieves a good tradeoff between the too-late and too-early \ac{HO} metrics. It significantly reduces the too-early \ac{HO} events (e.g., ping-pong \acp{HO}), without increasing the \ac{HO} failures.

%

\section{Conclusions}\label{sec:Conclusion}
In this paper, we proposed a novel \ac{SAMRO} solution with the newly introduced slice-specific \ac{HO} parameters and \acp{KPI}, optimizing both service quality and handover performance for diverse slices. We also proposed a two-step transfer learning-empowered actor-critic scheme that can deal with large discrete spaces, and enable safe and effective online training with regularized objective function and mixed replay buffer sampling. The system-level simulation shows significant performance benefits versus the baseline algorithms without slice-awareness. 
\acrodef{3GPP}{3rd generation partnership project}
\acrodef{5G}{fifth generation}

\acrodef{BS}{base station}
\acrodef{BCQ}{batch-constrained deep Q-learning}	
\acrodef{CQI}{channel quality indicator}
\acrodef{CIO}{cell individual offset}
\acrodef{CDF}{cumulative density function}
\acrodef{DEEN}{deep energy estimator network}
\acrodef{DNN}{deep neural network}
\acrodef{DDPG}{deep deterministic policy gradient}
\acrodef{DQN}{deep Q-learning}
\acrodef{DRL}{deep reinforcement learning}
\acrodef{eICIC}{enhanced inter-cell interference coordination}
\acrodef{eMBB}{enhanced mobile broadband}

\acrodef{GP}{Gaussian process}

\acrodef{HO}{handover}
\acrodef{HOM}{handover margin}
\acrodef{HOL}{Too-late-handover}
\acrodef{HOE}{Too-early-handover}
\acrodef{HOW}{Wrong-cell-handover}
\acrodef{HOPP}{Ping-pong-handover}
\acrodef{HFR}{handover failure ratio}
\acrodef{HetNet}{heterogeneous network}

\acrodef{ICI}{inter-cell interference}


\acrodef{KL}{Kullback-Leibler}
\acrodef{KPI}{Key Performance Indicator}
\acrodef{LTE}{long-term evolution} 
\acrodef{LSL}{latency service level}

\acrodef{MDP}{Markov decision process}
\acrodef{MRO}{mobility robustness optimization}
\acrodef{MLB}{mobility load balancing}
\acrodef{MLP}{multi-layer perception}

\acrodef{NLES}{nonlinear equation system}

\acrodef{OFDM}{orthogonal frequency division multiplexing}

\acrodef{PDF}{probability density function}

\acrodef{PPR}{ping-pong ratio}
    
\acrodef{QoE}{quality of experience}
\acrodef{QoS}{quality of service}

\acrodef{RAN}{radio access network}
\acrodef{RRM}{radio resource management}
\acrodef{RLF}{radio link failure}

\acrodef{SAC}{soft actor-critic}
\acrodef{SAMRO}{slice-aware mobility robustness optimization}
\acrodef{SNR}{signal-to-noise ratio}
\acrodef{SINR}{signal-to-interference-plus-noise ratio}
\acrodef{SIR}{signal-to-interference ratio}
\acrodef{SON}{self-organizing network}
\acrodef{SVM}{support vector machine}
\acrodef{SVD}{singular value decomposition}

 \acrodef{TTT}{time-to-trigger} 
\acrodef{TD3}{twin delayed deep deterministic policy gradient}
\acrodef{TSL}{throughput service level}

\acrodef{UE}{user equipment}
\acrodef{UI}{user interface}
\acrodef{URLLC}{ultra-reliable and low-latency communication}





\section*{Acknowledgement}
The authors would like to thank Senthil Kumaran K and Niraj Nanavaty for providing the practical insights.
\bibliographystyle{IEEEtran}
\bibliography{myreferences}

\end{document}